\documentclass[zpreprint,zbstnp,final]{zeus_paper}
\usepackage{placeins}
\usepackage[titletoc,title]{appendix}
\usepackage{authblk}
\usepackage{hyperref}
\usepackage{ifthen}
\usepackage{units}
\usepackage{rotating}
\usepackage{epsfig}
\usepackage{array}
\usepackage{multirow}
\usepackage{xspace}
\usepackage{upgreek}
\usepackage{pgfplots}
\usepackage{calc}
\usepackage{ifthen}
\usepackage{multirow}

\usepackage{caption}
\usepackage{listings}
\usepackage{tikz}
\renewcommand{\arraystretch}{1.00}

\definecolor{mygreen}{rgb}{0,0.6,0}
\definecolor{mygray}{rgb}{0.5,0.5,0.5}
\definecolor{mymauve}{rgb}{0.58,0,0.82}

\newcommand{\ypointsnomath}{\{0.1000,0.0215,0.0046,0.0010\}}
\newcommand{\NMIN}{2\xspace}
\newcommand{\NMAX}{4\xspace}

\newcommand{\NSUBSAMPLES}{ 1000\xspace}
\newcommand{\NEVENTS}{100000\xspace}
\newcommand{\NSTEP}{  100\xspace}
\newcommand{\ERRORSCALEFACTOR}{   50\xspace}
\newcommand{\BIGEXPRESSIONYR}{
\begin{pmatrix}
R_{2}(0.1000)\\
R_{2}(0.0215)\\
R_{2}(0.0046)\\
R_{2}(0.0010)\\
R_{3}(0.1000)\\
R_{3}(0.0215)\\
R_{3}(0.0046)\\
R_{3}(0.0010)\\
R_{4}(0.1000)\\
R_{4}(0.0215)\\
R_{4}(0.0046)\\
R_{4}(0.0010)\\
\end{pmatrix}=
\begin{pmatrix}
1&0&0&0&0&0&0&0&0&0&0&0&0&0&0\\
1&1&0&0&0&1&0&0&0&0&0&0&0&0&0\\
1&1&1&0&0&1&1&0&0&1&0&0&0&0&0\\
1&1&1&1&0&1&1&1&0&1&1&0&1&0&0\\
0&1&1&1&1&0&0&0&0&0&0&0&0&0&0\\
0&0&1&1&1&0&1&1&1&0&0&0&0&0&0\\
0&0&0&1&1&0&0&1&1&0&1&1&0&0&0\\
0&0&0&0&1&0&0&0&1&0&0&1&0&1&0\\
0&0&0&0&0&1&1&1&1&1&1&1&1&1&1\\
0&0&0&0&0&0&0&0&0&1&1&1&1&1&1\\
0&0&0&0&0&0&0&0&0&0&0&0&1&1&1\\
0&0&0&0&0&0&0&0&0&0&0&0&0&0&1\\
\end{pmatrix}
\begin{pmatrix}
\kappa_{2222}\\
\kappa_{2223}\\
\kappa_{2233}\\
\kappa_{2333}\\
\kappa_{3333}\\
\kappa_{2224}\\
\kappa_{2234}\\
\kappa_{2334}\\
\kappa_{3334}\\
\kappa_{2244}\\
\kappa_{2344}\\
\kappa_{3344}\\
\kappa_{2444}\\
\kappa_{3444}\\
\kappa_{4444}\\
\end{pmatrix}
}

\newcommand{\BIGEXPRESSIONYDall}{
\begin{pmatrix}
D_{23}[0.0215:0.1000]\\
D_{23}[0.0046:0.0215]\\
D_{23}[0.0010:0.0046]\\
D_{34}[0.0215:0.1000]\\
D_{34}[0.0046:0.0215]\\
D_{34}[0.0010:0.0046]\\
\end{pmatrix}=
\begin{pmatrix}
0&1&0&0&0&1&0&0&0&0&0&0&0&0&0\\
0&0&1&0&0&0&1&0&0&1&0&0&0&0&0\\
0&0&0&1&0&0&0&1&0&0&1&0&1&0&0\\
0&0&0&0&0&1&1&1&1&0&0&0&0&0&0\\
0&0&0&0&0&0&0&0&0&1&1&1&0&0&0\\
0&0&0&0&0&0&0&0&0&0&0&0&1&1&0\\
\end{pmatrix}
\begin{pmatrix}
\kappa_{2222}\\
\kappa_{2223}\\
\kappa_{2233}\\
\kappa_{2333}\\
\kappa_{3333}\\
\kappa_{2224}\\
\kappa_{2234}\\
\kappa_{2334}\\
\kappa_{3334}\\
\kappa_{2244}\\
\kappa_{2344}\\
\kappa_{3344}\\
\kappa_{2444}\\
\kappa_{3444}\\
\kappa_{4444}\\
\end{pmatrix}
}
\newcommand{\FISHER}{
$[-0.022,0.042]$ for $\rho=0.010$;
$[0.069,0.131]$ for $\rho=0.100$;
$[0.579,0.620]$ for $\rho=0.600$;
$[0.894,0.906]$ for $\rho=0.900$ and 
$[0.989,0.991]$ for $\rho=0.990$.
}

\newcommand{\NEVENTSPP}{250000\xspace}

\newcommand{\PTpointsnomath}{\{60,80,110,160,210\}\xspace}
\newcommand{\ERRORSCALEFACTORPT}{   10\xspace}
\newcommand{\BIGEXPRESSIONPT}{
\begin{pmatrix}
N_{1}[80:110]\\
N_{1}[110:160]\\
N_{1}[160:210]\\
N_{2}[60:80]\\
N_{2}[80:110]\\
N_{2}[110:160]\\
N_{2}[160:210]\\
N_{3}[60:80]\\
N_{3}[80:110]\\
N_{3}[110:160]\\
N_{3}[160:210]\\
\end{pmatrix}=
\begin{pmatrix}
1&0&0&0&1&0&0&0&0&0&0&0&0&0&1&0&0&0&1&0&0&0&0&0&0&0&0&0&1&0&0&0&0&0&0&0&0&0&0&0&0&0&0&0&0&0&0&0\\
0&1&0&0&0&1&0&0&1&0&0&0&0&0&0&1&0&0&0&1&0&0&1&0&0&0&0&0&0&1&0&0&1&0&0&0&0&0&1&0&0&0&0&0&0&0&0&0\\
0&0&1&0&0&0&1&0&0&1&0&1&0&0&0&0&1&0&0&0&1&0&0&1&0&1&0&0&0&0&1&0&0&1&0&1&0&0&0&1&0&1&0&0&1&0&0&0\\
1&1&1&1&0&0&0&0&0&0&0&0&0&0&1&1&1&1&0&0&0&0&0&0&0&0&0&0&0&0&0&0&0&0&0&0&0&0&0&0&0&0&0&0&0&0&0&0\\
0&0&0&0&1&1&1&1&0&0&0&0&0&0&0&0&0&0&1&1&1&1&0&0&0&0&0&0&1&1&1&1&0&0&0&0&0&0&0&0&0&0&0&0&0&0&0&0\\
0&0&0&0&0&0&0&0&1&1&1&0&0&0&0&0&0&0&0&0&0&0&1&1&1&0&0&0&0&0&0&0&1&1&1&0&0&0&1&1&1&0&0&0&0&0&0&0\\
0&0&0&0&0&0&0&0&0&0&0&1&1&0&0&0&0&0&0&0&0&0&0&0&0&1&1&0&0&0&0&0&0&0&0&1&1&0&0&0&0&1&1&0&1&1&0&0\\
0&0&0&0&0&0&0&0&0&0&0&0&0&0&1&1&1&1&1&1&1&1&1&1&1&1&1&1&0&0&0&0&0&0&0&0&0&0&0&0&0&0&0&0&0&0&0&0\\
0&0&0&0&0&0&0&0&0&0&0&0&0&0&0&0&0&0&0&0&0&0&0&0&0&0&0&0&1&1&1&1&1&1&1&1&1&1&0&0&0&0&0&0&0&0&0&0\\
0&0&0&0&0&0&0&0&0&0&0&0&0&0&0&0&0&0&0&0&0&0&0&0&0&0&0&0&0&0&0&0&0&0&0&0&0&0&1&1&1&1&1&1&0&0&0&0\\
0&0&0&0&0&0&0&0&0&0&0&0&0&0&0&0&0&0&0&0&0&0&0&0&0&0&0&0&0&0&0&0&0&0&0&0&0&0&0&0&0&0&0&0&1&1&1&0\\
\end{pmatrix}
\begin{pmatrix}
\kappa_{21000}\\
\kappa_{21100}\\
\kappa_{21110}\\
\kappa_{21111}\\
\kappa_{22000}\\
\kappa_{22100}\\
\kappa_{22110}\\
\kappa_{22111}\\
\kappa_{22200}\\
\kappa_{22210}\\
\kappa_{22211}\\
\kappa_{22220}\\
\kappa_{22221}\\
\kappa_{22222}\\
\kappa_{31000}\\
\kappa_{31100}\\
\kappa_{31110}\\
\kappa_{31111}\\
\kappa_{32000}\\
\kappa_{32100}\\
\kappa_{32110}\\
\kappa_{32111}\\
\kappa_{32200}\\
\kappa_{32210}\\
\kappa_{32211}\\
\kappa_{32220}\\
\kappa_{32221}\\
\kappa_{32222}\\
\kappa_{33000}\\
\kappa_{33100}\\
\kappa_{33110}\\
\kappa_{33111}\\
\kappa_{33200}\\
\kappa_{33210}\\
\kappa_{33211}\\
\kappa_{33220}\\
\kappa_{33221}\\
\kappa_{33222}\\
\kappa_{33300}\\
\kappa_{33310}\\
\kappa_{33311}\\
\kappa_{33320}\\
\kappa_{33321}\\
\kappa_{33322}\\
\kappa_{33330}\\
\kappa_{33331}\\
\kappa_{33332}\\
\kappa_{33333}\\
\end{pmatrix}
}

\newcommand{\NEVENTSEP}{500000\xspace}

\newcommand{\ERRORSCALEFACTORET}{    5\xspace}
\newcommand{\BIGEXPRESSIONET}{
\begin{pmatrix}
N({\cal R}=0.5)\\
N({\cal R}=0.7)\\
N({\cal R}=1.0)\\
\end{pmatrix}=
\begin{pmatrix}
0&0&0&0&0&0&0&0&0&0&0&0&0&0&0&1&1&1&1&1&1&1&1&1&1&1&1&1&1&1&1&2&2&2&2&2&2&2&2&2&2&2&2&2&2&2&2&3&3&3&3&3&3&3&3&3&3&3&3&3&3&3&3\\
0&0&0&1&1&1&1&2&2&2&2&3&3&3&3&0&0&0&0&1&1&1&1&2&2&2&2&3&3&3&3&0&0&0&0&1&1&1&1&2&2&2&2&3&3&3&3&0&0&0&0&1&1&1&1&2&2&2&2&3&3&3&3\\
1&2&3&0&1&2&3&0&1&2&3&0&1&2&3&0&1&2&3&0&1&2&3&0&1&2&3&0&1&2&3&0&1&2&3&0&1&2&3&0&1&2&3&0&1&2&3&0&1&2&3&0&1&2&3&0&1&2&3&0&1&2&3\\
\end{pmatrix}
\begin{pmatrix}
\kappa_{001}\\
\kappa_{002}\\
\kappa_{003}\\
\kappa_{010}\\
\kappa_{011}\\
\kappa_{012}\\
\kappa_{013}\\
\kappa_{020}\\
\kappa_{021}\\
\kappa_{022}\\
\kappa_{023}\\
\kappa_{030}\\
\kappa_{031}\\
\kappa_{032}\\
\kappa_{033}\\
\kappa_{100}\\
\kappa_{101}\\
\kappa_{102}\\
\kappa_{103}\\
\kappa_{110}\\
\kappa_{111}\\
\kappa_{112}\\
\kappa_{113}\\
\kappa_{120}\\
\kappa_{121}\\
\kappa_{122}\\
\kappa_{123}\\
\kappa_{130}\\
\kappa_{131}\\
\kappa_{132}\\
\kappa_{133}\\
\kappa_{200}\\
\kappa_{201}\\
\kappa_{202}\\
\kappa_{203}\\
\kappa_{210}\\
\kappa_{211}\\
\kappa_{212}\\
\kappa_{213}\\
\kappa_{220}\\
\kappa_{221}\\
\kappa_{222}\\
\kappa_{223}\\
\kappa_{230}\\
\kappa_{231}\\
\kappa_{232}\\
\kappa_{233}\\
\kappa_{300}\\
\kappa_{301}\\
\kappa_{302}\\
\kappa_{303}\\
\kappa_{310}\\
\kappa_{311}\\
\kappa_{312}\\
\kappa_{313}\\
\kappa_{320}\\
\kappa_{321}\\
\kappa_{322}\\
\kappa_{323}\\
\kappa_{330}\\
\kappa_{331}\\
\kappa_{332}\\
\kappa_{333}\\
\end{pmatrix}
}

\newcommand{\eVdist}{\kern-0.06667em}
\newcommand{\GeV}{{\,\text{Ge}\eVdist\text{V\/}}}
\newcommand{\TeV}{{\,\text{Te}\eVdist\text{V\/}}}

\newcommand{\epjcbreak}[1]{}
\newcommand{\nimbreak}[1]{}
\newcommand{\jinstbreak}[1]{}
\newcommand{\draftbreak}[1]{}
\newcommand{\arxivbreak}[1]{\\#1}

\newcommand{\FASTJET}{FastJet\,3.1.3\xspace}
\newcommand{\ppsherpa}{SHERPA\,2.2.1\xspace}
\newcommand{\eesherpa}{SHERPA\,2.2.1\xspace}
\newcommand{\epsherpa}{SHERPA\,2.2.1\xspace}

\newcommand{\tabfontsize}{\small}

\title{Studies of correlations between measurements of jet observables}

\author[a]{Andrii Verbytskyi\thanks{andrii.verbytskyi@mpp.mpg.de}}

\affil[a]{Max-Planck-Institut f\"{u}r Physik (Werner-Heisenberg-Institut),\protect\\ F\"{o}ringer Ring 6, M\"unchen 80805, Germany }

\prepnum{MPP-2016-300}

\abstract{We present a  method for calculation of statistical correlations between 
 measured jet observables in high energy  collisions. The 
 case of measurements of jet rates in $e^+e^-$ collisions is considered 
 in detail.  The 
  method is compared to sampling based methods used in the past.
 
}

\begin{document}
\makezeustitle\newpage
\clearpage
\pagenumbering{arabic}
\pagestyle{plain}
\newpage
\section{Introduction}
\label{sec:int}
In high energy particle collisions the partons and hadrons are produced 
in collimated bunches called {\it jets}. In a very simplified model a 
jet can be considered as a parton that was produced in a hard process 
and went through showering and hadronisation processes. The studies of 
jet  production in the $e^+e^-$, $e^{\pm}p$ and $pp$ collisions serve 
as a strong test of Quantum Chromodynamics (QCD). 
 For instance, 
 comparisons of the jet production cross sections in $e^+e^-$ 
collisions to the fixed order or resummed perturbative QCD predictions 
are used for  the extraction of the  strong coupling constant 
$\alpha_s(M_Z)$~\cite{Dissertori:2009qa,Schieck:2012mp,OPAL:2011aa,
Zomorrodian:2015dom}.
For the ultimate precision of the $\alpha_s(M_Z)$  determination, 
several measured jet observables  are combined. In this case the 
statistical and systematical correlations between the measurements of 
jet observables  are important.  To estimate the statistical 
correlations between the  measured quantities,  
several methods were used in the past.

In this work  jet observables {\it classes} are introduced and are used to  calculate other jet observables 
and  obtain statistical correlations between them in a model-independent
 way  using only data.  To illustrate the method,  calculations of  jet rates in the $e^+e^-$ collisions,
 multijet cross sections in $pp$ collisions and inclusive jet cross sections in $e^{\pm}p$ collisions
are considered in  toy analyses with Monte Carlo (MC) simulated events.  
The results for  jet rates in $e^+e^-$ collisions are compared to those obtained with sampling method used in the 
 past~\cite{Heister:2003aj,OPAL:2011aa,
 Schieck:2012mp,Zomorrodian:2015dom}.
\section{Jet algorithms}
\label{sec:jets}
A jet clustering algorithm is a way to simplify high energy collision 
event topology and exhibit the underlying physics at the parton level. 
The main goal of such a procedure is to reconstruct the kinematic 
variables of the partons produced during the primary hard interaction. 
The energy and the momenta of the partons are reconstructed by combining
  momenta and energy of the  charged and neutral particles which are 
  clustered into jets. Several jet algorithms are used to perform the 
  combination in different  environments --  $e^+e^-$,  $pp$ or 
  $e^{\pm}p$ collisions.  A detailed overview of their properties can 
  be found elsewhere~\cite{Ali:2010tw,SchornerSadenius:2012de}, only some properties are briefly discussed below.
  
The number of jets reconstructed a given collision event depends 
on the applied jet algorithm and used {\it cut parameters}. 
The former can be illustrated with 
 the  $k_T$ algorithm~\cite{Catani:1991hj}
 widely used in $e^+e^-$ jet studies. 
In this algorithm a
measure $y_{ij}$ is defined for a pair of particles $i$ and $j$
with total visible energy in the event $E_{\rm vis}$, energies
 of particles $E_i$, $E_j$ and angle between them $\theta_{ij}$ 
\begin{equation}
y_{ij}=\frac{2min(E_i^2,E_j^2)(1-\cos{\theta_{ij}})}{E_{\rm vis}^2}.
\end{equation}
At every step of the iterative recombination procedure, all distances 
are calculated and a pair 
of particles with minimal distance between them is determined.
As 
 long as this distance is smaller than cut parameter $y$, these 
 particles  are combined by adding their $4$-vectors and next iteration
  begins. In this 
 way $y$ defines the number of jets in the event. 
 
To define the number of jets obtained with  cone-like 
 algorithms with a minimum energy 
 requirement for a jet, the minimal half opening angle of jet 
 cone ${\cal R}$  can be used.
The  anti-$k_T$~\cite{Cacciari:2008gp} or SISCone~\cite{Salam:2007xv} 
 can serve as examples of such cone-like jet algorithms.
For the inclusive clustering algorithms, often used in $e^{\pm}p$ and 
$pp$ jet studies, or for the cone-like algorithms~\cite{Cacciari:2008gp,Salam:2007xv} with fixed ${\cal R}$ 
parameter, the number of hard jets is defined with a cut on jet 
(transverse) momentum or jet (transverse) 
energy~\cite{SchornerSadenius:2012de}.

The quantities of interest measured in $e^+e^-$ jet 
studies~\cite{Heister:2003aj,OPAL:2011aa,Schieck:2012mp,
Zomorrodian:2015dom}
are ${\cal N}$-jet rates and the distribution of transition parameters 
$D_{i,i+1}$. The ${\cal N}$-jet rates are defined as a ratio of 
production cross section of events with ${\cal N}$ jets at some $y$ 
value to the total hadronic cross section. The transition parameters 
$D_{i,i+1}$ are defined as values of $y$ for which the event changes 
from $i$ to $i+1$ jets.

In the $e^{\pm}p$ jet studies the quantities of 
interest are the multi-jet  cross sections in bins  
of transverse momentum (energy) of jets and  photon 
virtuality~\cite{Chekanov:2005ve,Andreev:2014wwa} and inclusive jet cross sections~\cite{Chekanov:2006yc}.
Similarly, in the $pp$ jet studies the multi-jet 
differential cross sections are measured in bins of transverse  
momentum of leading jets~\cite{Aad:2011tqa}.
\section{Event classes and relation to other jet observables}
\label{sec:class}
\subsection{Event classes}
\begin{figure}[!htbp]\centering
\resizebox{1.0\linewidth}{0.58\linewidth}{
{\bf \boldmath\input{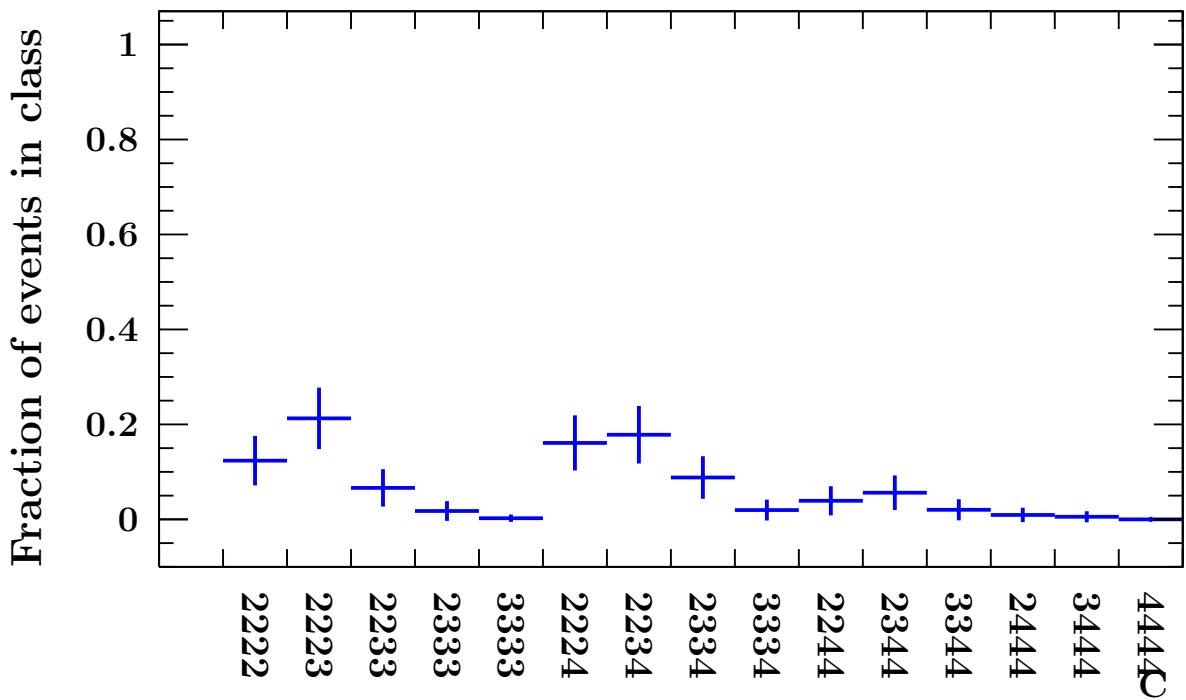}}
}
\caption{Fractions of event classes obtained 
with the exclusive  $k_{T}$ algorithm at particle level
from  $e^+e^-$
sample simulated with
\eesherpa using $\alpha_s(M_Z)=0.12$.
 The vertical error bars show the square 
 roots of the covariance matrix diagonal 
 elements  multiplied by  factor \ERRORSCALEFACTOR for better visibility. 
 }
\label{fig:classes}
\end{figure}

An application of jet algorithm with a fixed set of cut
parameters $\{y_1 <y_{2}<\dots <y_n\}$ results in  set  
$C={\{ r_{1} r_{2}\dots r_{n}\}}$ of natural numbers for each 
event, where $r_{i}$ corresponds to number of jets obtained with 
cut parameter $y_{i}$.   As events with same $C$ have the same topology, 
$C$ is used to group events in 
{\it classes}.
To limit the number of possible classes,  events with number 
of jets ${\cal N}<{\cal N}_{min}$  are treated  as events with ${\cal N}_{min}$ jets 
and events with number of jets ${\cal N}>{\cal N}_{max}$ as events with 
${\cal N}_{max}$ jets. In the general case the total number of classes, 
$K$, is~\footnote{For each $y_{i}$ the number of jets can be between 
${\cal N}_{min}$ and ${\cal N}_{max}$. Hence, $K$ is a number of 
combinations of ${\cal N}_{max}-{\cal N}_{min}+1$ distinct values  by 
$n$ with repetitions.
  }
\begin{equation}\label{eq:ngeneral}
K=({\cal N}_{max}-{\cal N}_{min}+1)^n.
\end{equation}
For the exclusive $k_T$ algorithm  in $e^+e^-$ collisions, the number of
 jets rises monotonically with the decreasing cut parameter $y$
and the number of classes is smaller~\footnote{Let us add 
$y_{0}$ to the set of cut parameters and  set $r_0=N_{max}-r_1$. 
With this addition, every $r_{i}\ge0$ and their sum is 
${\cal N}_{max}-{\cal N}_{min}$. Hence, $K$ is a number of weak 
compositions of ${\cal N}_{max}-{\cal N}_{min}$  by $n+1$.
}:
\begin{equation}
K=\frac{(n+{\cal N}_{max}-{\cal N}_{min})!}{n!({\cal N}_{max}-
{\cal N}_{min})!}.
\end{equation}
An example of event classes  distribution in $e^+e^-$ collisions is given in Fig.~\ref{fig:classes}. 

\FloatBarrier

\begin{figure}[!htbp]\centering
\resizebox{1.0\linewidth}{0.62\linewidth}{
{\bf \boldmath\input{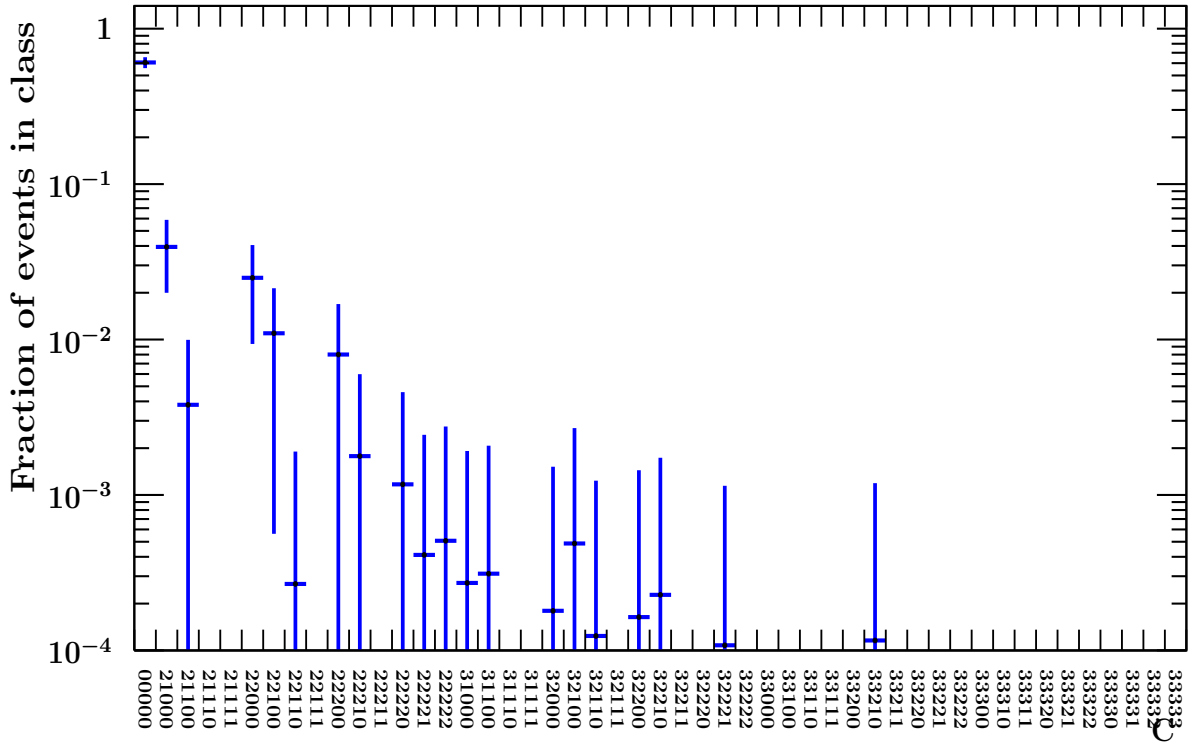}}
}
\caption{Distribution of event classes obtained 
with the anti-$k_{T}$ algorithm at particle level
from  $pp$
sample simulated with
\eesherpa.
 The vertical error bars show the square 
 roots of the covariance matrix diagonal 
 elements multiplied by factor \ERRORSCALEFACTORPT for better visibility. 
 Bin labelled with $00000$ contains events rejected with selection cuts.
 }
\label{fig:ppclasses}
\end{figure}

The sample contains \NEVENTS events  simulated with the \eesherpa 
MC generator~\cite{Gleisberg:2008ta} at centre-of-mass-energy $91\GeV$ (see Listing~\ref{lst:eesherpa} for details).
The exclusive $k_T$ algorithm as implemented in \FASTJET~\cite{Cacciari:2011ma} is applied at particle level. The used 
cut parameters are  subset of these in  Ref.~\cite{OPAL:2011aa}
\begin{equation}
\ypointsnomath 
\label{YPOINTS}
\end{equation}
  and ${\cal N}_{min}=\NMIN$, 
${\cal N}_{max}=\NMAX$.
\FloatBarrier
Similarly, for the anti-$k_T$ algorithm~\cite{Cacciari:2008gp}  in $pp$ collisions, 
the number of jets rises monotonically with the decreasing $p_{T}$ cut, which acts as the cut parameter in this case.
An example of event classes  distribution in $pp$ collisions is given 
in Fig.~\ref{fig:ppclasses}. 

The sample contains \NEVENTSPP events  simulated with the \ppsherpa MC generator~\cite{Gleisberg:2008ta} at centre-of-mass-energy $7\TeV$ (see Listing~\ref{lst:ppsherpa} for details).
The  anti-$k_{T}$ jet algorithm from \FASTJET~\cite{Cacciari:2011ma}  with the  radius parameter ${\cal R}=0.4$  is used at particle level. Similarly to Ref.~\cite{Aad:2011tqa}, 
only jets with transverse momenta $p_{T}>60\GeV$ and rapidity $|y|<2.8$ are considered.
The leading jet is further required to have $p_{T}>80\GeV$.
The used  $p_{T}$ cuts  are
\begin{equation}
\PTpointsnomath\GeV,
\label{PTPOINTS}
\end{equation} and ${\cal N}_{min}=0$,  ${\cal N}_{max}=3$.

\begin{figure}[!htbp]\centering
\resizebox{1.0\linewidth}{0.62\linewidth}{
{\bf \boldmath\input{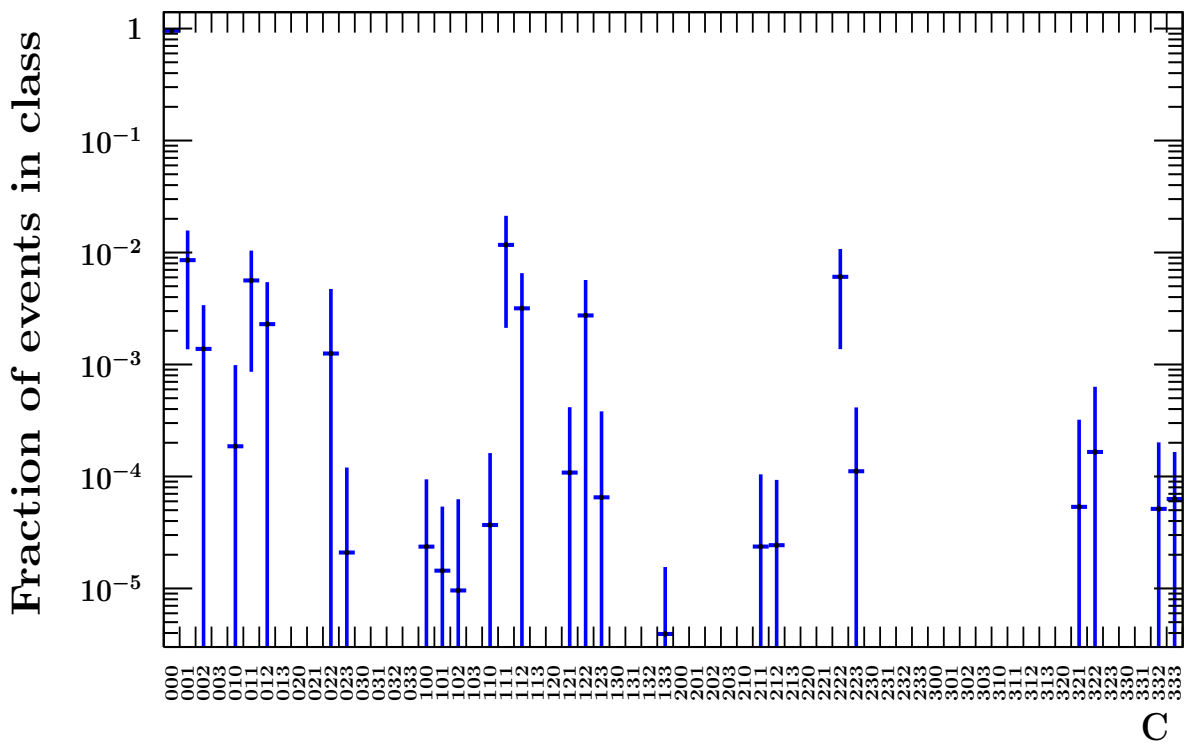}}
}
\caption{Distribution of event classes obtained 
with the $k_{T}$ algorithm at particle level
from  $e^{-}p$
sample simulated with
\epsherpa.
 The vertical error bars show the square 
 roots of the covariance matrix diagonal 
 elements multiplied by  factor \ERRORSCALEFACTORET for better visibility. 
 Bin labelled with $000$ contains events rejected with selection cuts.
 }
\label{fig:epclasses}
\end{figure}

\FloatBarrier
The jet radius can act as a cut parameter as well. However, the number of jets does not change 
 monotonically with the change of radius parameter and the total number of classes follows Eq.~\ref{eq:ngeneral}.
An example of usage of jet radius as a cut parameter can be found in the jet studies in deep-inelastic scattering in $e^{\pm}p$ 
collisions~\cite{Chekanov:2006yc}.
An example of event classes distribution for this case 
 is given in Fig.~\ref{fig:epclasses}. 

The sample contains \NEVENTSEP weighted events  simulated with the 
\epsherpa 
MC generator~\cite{Gleisberg:2008ta} at centre-of-mass-energy $318\GeV$.
The $k_{T}$ jet algorithm is used at particle level in the Breit frame. 
The implementation from \FASTJET~\cite{Cacciari:2011ma} is modified to calculate 
distances between jets with transverse energy as in Ref.~\cite{Chekanov:2006yc}.
The selected 
kinematic range is $125<Q^{2}<5000\GeV^{2}$ and  $|\cos\gamma_{h}|<0.65$,  where $Q^{2}=-|\vec{q}|^{2}$ with $\vec{q}$  standing for the  four momentum 
of the  exchanged 
boson and $\gamma_{h}$ is equivalent to the polar angle of the scattered quark in the leading-order quark-parton model~\cite{Bentvelsen:1992fu}.
Only jets with transverse energy $E_{T}^{B}>8\GeV$ and pseudorapidity $-2<\eta^{B}<1.5$  in Breit frame 
and  $E_{T}^{L}>2.5\GeV$ in laboratory frame  
are considered.
 Events contained jets with  pseudorapidity  $\eta^{L}<-2$ in the laboratory frame are removed.
The used jet radius parameters are 
\begin{equation}
\{0.5,0.7,1.0\}
\label{ETPOINTS}
\end{equation} and ${\cal N}_{min}=0$, ${\cal N}_{max}=3$.
\FloatBarrier
\subsection{Relation to jet rates in $e^{+}e^{-}$ collisions}
\label{sec:class:jr}
\begin{figure}[!htbp]\centering
\resizebox{1.0\linewidth}{0.62\linewidth}{
{\bf \boldmath\input{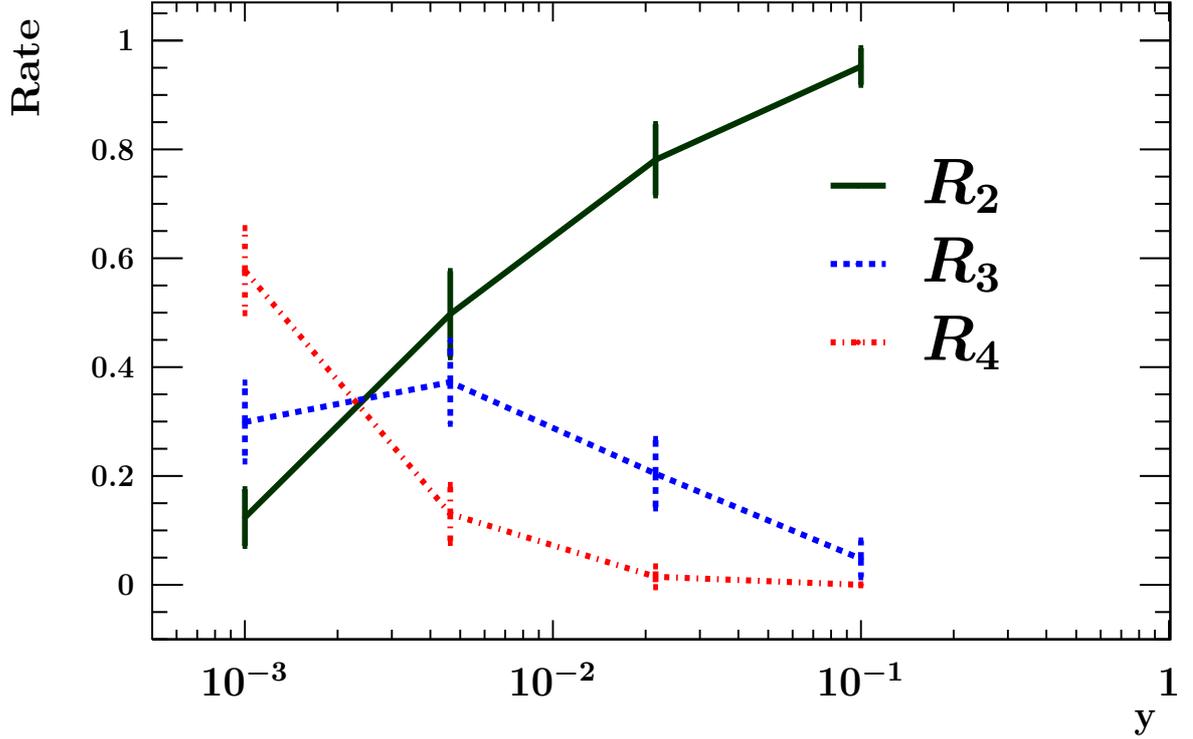}}}
\caption{Jet rates obtained 
with the exclusive  $k_{T}$ algorithm at particle level
from $e^+e^-$
sample simulated with 
\eesherpa using $\alpha_s(M_Z)=0.12$. 
 The vertical error bars show the square roots of the covariance
  matrix diagonal 
 elements. For better visibility the size of error bars is scaled by  factor 
 \ERRORSCALEFACTOR  and individual points are connected with straight lines.
 }
\label{fig:rates}
\end{figure}

The fraction of classes in $e^{+}e^{-}$ collisions can be related to ${\cal N}$-jet rates and 
transition parameters.
The fraction of events of a given class 
is denoted with $\kappa_{r_{1}r_{2}\dots r_n}$ or 
$\kappa_{i=1\dots K}$.
 The ${\cal N}$-jet rates for every $y$ value, 
$R_{{\cal N}={\cal N}_{min}\dots {\cal N}_{max}}(y=y_1y_2\dots y_n)$,  
can be  obtained from the fraction of events in classes with
\begin{equation}
R_{{\cal N}}(y_i)=\sum_{\{r_i=N\}}\kappa_{1\dots K}.
\end{equation}
For instance, for the chosen set of cut  parameters from 
Eq.~\ref{YPOINTS} for the exclusive $k_T$ algorithm 
\begin{equation}
R_{3}(y_2)=\kappa_{2333}+\kappa_{3333}+\kappa_{3334}+\kappa_{3344}.
\end{equation}
In a general case for all jet rates this can be expressed  as a 
linear transformation from fraction of event classes:
\begin{equation}
\overrightarrow{(R_{{\cal N}={\cal N}_{min}\dots {\cal N}_{max}}
(y=y_1y_2\dots y_n))}=A_R 
\overrightarrow{(\kappa_{1\dots K})},
\end{equation}
where the matrix  $A_R$ is constructed to sum all the contributions from
 all classes to the particular 
 $R_{{\cal N}={\cal N}_{min}\dots {\cal N}_{max}}(y=y_1y_2\dots y_n)$. 
This relation by construct delivers results identical to a simple jet 
counting. In the particular case of the chosen set of cut 
parameters
\renewcommand*{\arraystretch}{0.65}
\begin{equation}
\arraycolsep=1pt
\BIGEXPRESSIONYR
.
\label{YR}
\end{equation}
The jet rates obtained with Eq.~\ref{YR} from the distribution in 
Fig.~\ref{fig:classes} are shown in Fig.~\ref{fig:rates}.

 \subsection{Relation to differential jet rates in $e^{+}e^{-}$ collisions}
 The fraction of events
with transition parameters in bins $[y_j:y_{j+1}]$ can be obtained 
similarly to the jet rates:
\begin{equation}
\overrightarrow{(D_{i,i+1}[y_j:y_{j+1}],j<n-1,i={\cal N}_{min}
\dots{\cal N}_{max}-1})=A_D 
\overrightarrow{(\kappa_{1\dots K})}.
\end{equation}
The differential-jet-multiplicity distributions~\cite{Komamiya:1989hw} 
can be obtained from $D_{i,i+1}[y_j:y_{j+1}]$ distributions with a 
division by corresponding bin width $y_j-y_{j+1}$. In the particular 
case of the chosen set of cut parameters
\renewcommand*{\arraystretch}{0.65}
\begin{equation}
\arraycolsep=1pt
\BIGEXPRESSIONYDall
.
\label{YD}
\end{equation}
\subsection{Relation to multijet cross sections in $pp$ collisions}
\begin{figure}[!htbp]\centering
\resizebox{1.0\linewidth}{0.62\linewidth}{
{\bf \boldmath\input{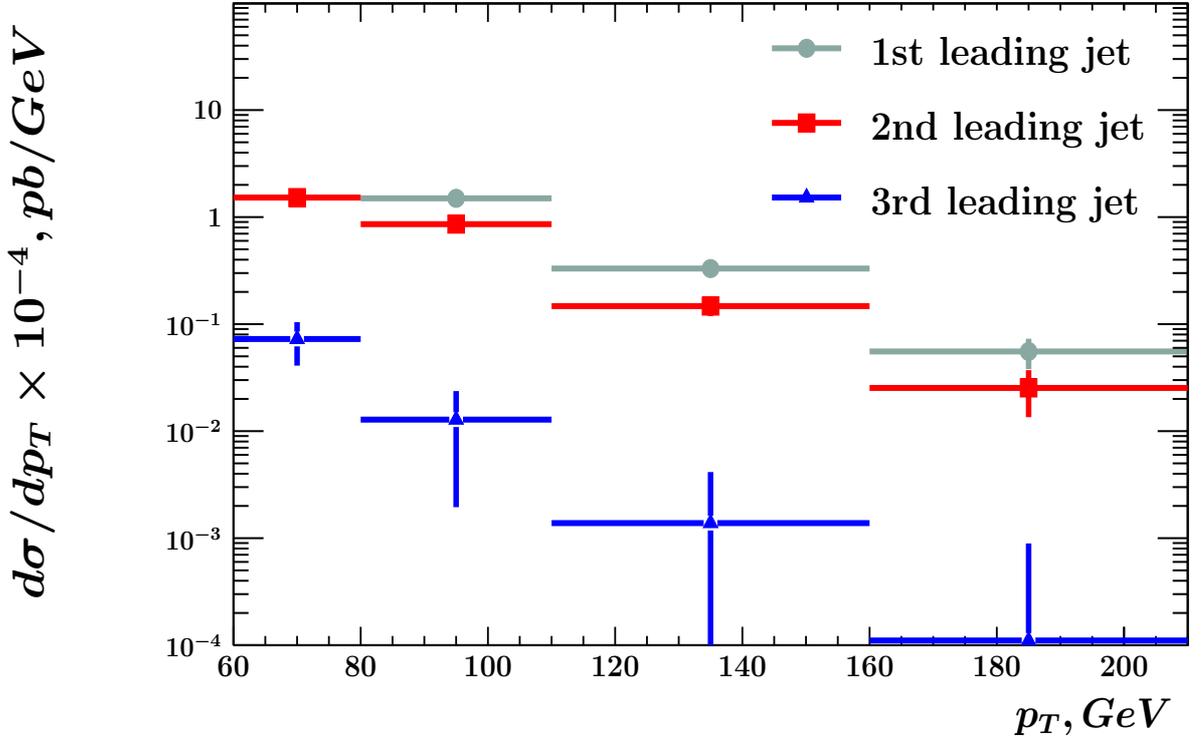}}}
\caption{Multijet cross sections obtained 
with the anti-$k_{T}$ algorithm at particle level
from $pp$
sample simulated with 
\ppsherpa. 
 The vertical error bars show the square roots of the covariance
  matrix diagonal 
 elements. For  better visibility the size of error bars is scaled by  factor 
 \ERRORSCALEFACTORPT.
 }
\label{fig:pprates}
\end{figure}

The distribution of classes in $pp$ collisions, obtained above, can be related to 
multijet cross sections.
Similarly to the $e^{+}e^{-}$ case, a matrix $A_{L J}$ can be constructed to give
\begin{equation}
\overrightarrow{(N_{i}[p_{T,j}:p_{T,j+1}],i=1
\dots{\cal N}_{max},j=1\dots n-1})=A_{L J} 
\overrightarrow{(\kappa_{1\dots K})}, 
\end{equation}
where $N_{i}[p_{T,j}:p_{T,j+1}]$ stands for the number of events with $i$-th leading jet in bin
$[p_{T,j}:p_{T,j+1}]$.
A division of numbers $N_{i}[p_{T,j}:p_{T,j+1}]$  by the  
bin widths $p_{T,j+1}-p_{T,j}$ an luminosity of the simulated sample will deliver  jet cross sections in 
corresponding bins.
In the particular 
case of the chosen set of $p_{T}$ cuts and additional requirements~\cite{Aad:2011tqa} we have:

\renewcommand*{\arraystretch}{0.7}
\begin{equation}
\arraycolsep=0.7pt
{\tiny
\BIGEXPRESSIONPT
},
\label{RP}
\end{equation}
where the boundaries of $p_{T}$ bin are given in $\GeV$.  The multijet cross sections obtained in this way from the distribution in  Fig.~\ref{fig:ppclasses} are shown in Fig.~\ref{fig:pprates}.

\FloatBarrier
\FloatBarrier
\subsection{Relation to inclusive jet cross sections in $e^{\pm}p$ collisions}
The distribution of classes in $e^{-}p$ collisions, obtained above, can be related to 
inclusive jet cross sections. The corresponding matrix $A_{I J}$ can be constructed to give
\begin{equation}
\overrightarrow{(N({\cal R}={\cal R}_{i}),i=1\dots n})=A_{I J}\overrightarrow{(\kappa_{1\dots K})}, 
\end{equation}
where $N({\cal R}={\cal R}_{i})$ stands for the number of inclusive jets obtained with radius parameter ${\cal R}_{i}$.
Unlike $A_{L J}$, the $A_{I J}$  matrix  contains coefficients greater than one, i.e.\  
$2$ for events with two jets and $3$ for events with three jets.
In the particular  case of chosen radius parameters
\renewcommand*{\arraystretch}{0.7}
\begin{equation}
\arraycolsep=0.6pt
{\tiny
\BIGEXPRESSIONET
},
\label{RE}
\end{equation}

\begin{figure}[!htbp]\centering
\resizebox{1.0\linewidth}{0.62\linewidth}{
{\bf \boldmath\input{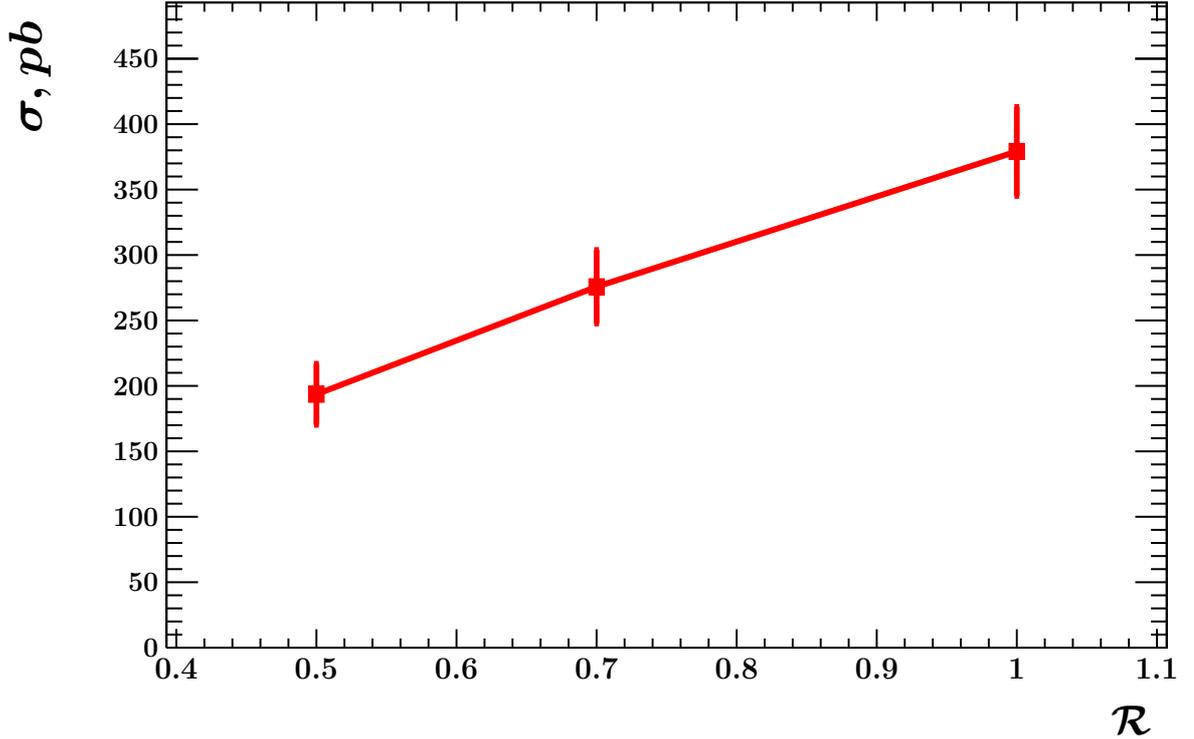}}}
\caption{Inclusive jet cross sections obtained 
with the $k_{T}$ algorithm at particle level
from $e^{-}p$
sample simulated with 
\epsherpa. 
 The vertical error bars show the square roots of the covariance
  matrix diagonal 
 elements. 
For better visibility the size of error bars is scaled by  factor 
 \ERRORSCALEFACTORET  and individual points are connected with straight lines.
 }
\label{fig:eprates}
\end{figure}

A division of the  $N({\cal R}={\cal R}_{i})$ numbers  by the luminosity of the simulated 
sample  delivers inclusive jet cross sections shown in Fig.~\ref{fig:eprates}.
\section{Statistical correlations}
In this section different methods to calculate the
statistical correlations between jet observables are described.
As every experimental measurement has to be corrected for instrumental 
(detector) effects, particular interest has the propagation of these 
corrections within every method.

\subsection{Classes-bases method}
In the classes-based method every event is  assigned to a class 
uniquely and the  number of events  in every class (e.g.\ at detector 
level in data or at particle level in MC simulated events) 
are statistically independent. 
Hereby, in the discussed cases of $pp$ and $e^{\pm}p$ collisions, the number of events in every class, $\kappa_{i}$, 
the  corresponding  covariance matrices are diagonal
\begin{equation}
 V^{\kappa}_{ij}=\delta_{ij}\kappa_{i}.
\end{equation}
The covariance matrix for multijet cross sections, $V^{L J}$, can  be obtained from the 
$V^{\kappa}$ using the matrix from Eq.~\ref{RP}:
\begin{equation}
 V^{L J}=A_{L J} V^{\kappa}A_{L J}^T.
\label{eq:ppcov}
\end{equation}
The covariance matrix for inclusive jet cross sections, $V^{I J}$, can  be obtained from the 
$V^{\kappa}$ using the matrix from Eq.~\ref{RE}:
\begin{equation}
 V^{I J}=A_{I J} V^{\kappa}A_{I J}^T.
\label{eq:epcov}
\end{equation}
 The numerical values for corresponding correlation matrices   are
  given in Tab.~\ref{tab:correlationsmcpp} and Tab.~\ref{tab:correlationsmcep}. 
\setlength{\tabcolsep}{1pt}
\renewcommand{\arraystretch}{1.25}
\begin{table}[!htbp]\centering\tabfontsize
\resizebox{\textwidth}{!}{\begin{tabular}{|ccc|cccc|cccc||cc|}\hline
\multicolumn{3}{|c}{$\frac{d\sigma}{dp_{T}}$, 1st leading jet }&\multicolumn{4}{|c}{$\frac{d\sigma}{dp_{T}}$, 2nd leading jet }&\multicolumn{4}{|c||}{$\frac{d\sigma}{dp_{T}}$, 3rd leading jet }& &\\
{\tiny $[80:110]$}&{\tiny $[110:160]$}&{\tiny $[160:210]$}&{\tiny $[60:80]$}&{\tiny $[80:110]$}&{\tiny $[110:160]$}&{\tiny $[160:210]$}&{\tiny $[60:80]$}&{\tiny $[80:110]$}&{\tiny $[110:160]$}&{\tiny $[160:210]$}&&\\\hline\hline
+1.00 &-0.04 &-0.02 &+0.73 &+0.49 &-0.03 &-0.01 &+0.03 &-0.00 &-0.00 &-0.00 &{\tiny $[80:110]$}&\multirow{3}{*}{$\frac{d\sigma}{dp_{T}}$, 1st leading jet}\\
    &+1.00 &-0.01 &+0.10 &+0.37 &+0.51 &-0.01 &+0.13 &+0.03 &+0.00 &-0.00 &{\tiny $[110:160]$}&\\
    &    &+1.00 &-0.00 &+0.03 &+0.33 &+0.46 &+0.16 &+0.14 &+0.05 &-0.00 &{\tiny $[160:210]$}&\\\hline
    &    &    &+1.00 &-0.04 &-0.02 &-0.01 &+0.06 &-0.01 &-0.00 &-0.00 &{\tiny $[60:80]$}&\multirow{4}{*}{$\frac{d\sigma}{dp_{T}}$, 2nd leading jet}\\
    &    &    &    &+1.00 &-0.02 &-0.01 &+0.08 &+0.04 &-0.00 &-0.00 &{\tiny $[80:110]$}&\\
    &    &    &    &    &+1.00 &-0.00 &+0.09 &+0.10 &+0.07 &-0.00 &{\tiny $[110:160]$}&\\
    &    &    &    &    &    &+1.00 &+0.08 &+0.06 &+0.04 &+0.03 &{\tiny $[160:210]$}&\\\hline
    &    &    &    &    &    &    &+1.00 &-0.00 &-0.00 &-0.00 &{\tiny $[60:80]$}&\multirow{4}{*}{$\frac{d\sigma}{dp_{T}}$, 3rd leading jet}\\
    &    &    &    &    &    &    &    &+1.00 &-0.00 &-0.00 &{\tiny $[80:110]$}&\\
    &    &    &    &    &    &    &    &    &+1.00 &-0.00 &{\tiny $[110:160]$}&\\
    &    &    &    &    &    &    &    &    &    &+1.00 &{\tiny $[160:210]$}&\\\hline
\end{tabular}
}
\caption{The  correlation matrix of   multijet cross sections in $pp$  collisions in bins of jet transverse momenta (given in brackets) 
calculated  with classes-based method. 
}
\label{tab:correlationsmcpp}
\end{table}
\setlength{\tabcolsep}{1pt}
\renewcommand{\arraystretch}{1.25}
\begin{table}[!htbp]\centering\tabfontsize
\begin{tabular}{|ccc||c|}\hline
$\sigma({\cal R}=0.5)$&$\sigma({\cal R}=0.7)$&$\sigma({\cal R}=1.0)$&\\\hline\hline
+1.00&+0.87&+0.76&$\sigma({\cal R}=0.5)$\\
    &+1.00&+0.89&$\sigma({\cal R}=0.7)$\\
    &    &+1.00&$\sigma({\cal R}=1.0)$\\
\hline\end{tabular}

\caption{The  correlation matrix of inclusive jet cross sections in $e^{-}p$ collisions obtained with different radius parameter ${\cal R}$ 
calculated  with classes-based method. 
}
\label{tab:correlationsmcep}
\end{table}

In the discussed case of $e^{+}e^{-}$ collisions, it is more practical to use 
the fraction of events in every class $\kappa_{i}$. 
It follows a multinomial distribution with the corresponding  covariance matrix 
\begin{equation}
 V^{\kappa}_{ij}=\kappa_{i}(\delta_{ij}-
 \kappa_{j}).
\end{equation}
Therefore, covariance matrices for the jet rates $V^R$ and transition 
parameters $V^D$ can be obtained from the 
$V^{\kappa}$ using the matrices from Eq.~\ref{YR} and Eq.~\ref{YD}:
\begin{equation}
 V^{R}=A_R V^{\kappa}A_R^T,  \ 
V^{D}=A_D V^{\kappa}A_D^T.
\label{eq:cov}
\end{equation}
The remaining part of 
this section will be dedicated to descriptions and comparisons to other 
methods.
\subsection{Direct counting method}
\label{sec:dircount}
In  the recent studies of jet production in $e^{\pm}p$ 
collisions~\cite{Lontkovskyi:2015} the problem of the estimation of 
statistical correlations between measured jet observables is addressed.
Namely, the direct counting method~\cite{Lontkovskyi:2015} is used to 
estimate  correlations between the measured jet cross sections in 
bins of transverse energy. The method implies counting of the events 
which contribute to different bins assuming
Poisson distribution for the number of events contributing in 
different ways. 
The obtained covariance matrices should be corrected for detector 
effects.
In the recent 
measurements of multijet jet production in $e^{\pm}p$ 
collisions~\cite{Andreev:2014wwa,Andreev:2016tgi}  the covariance 
matrices at  particle level are obtained with regularised unfolding 
method (see Refs.~\cite{Adye:2011gm,Schmitt:2012kp} for details of 
implementation). 
The procedure is straightforward and  well established, but depends on 
the regularisation conditions and can be 
problematic in case of singular covariance  matrix at detector level
(e.g.\ because of small number of data events).
\subsection{Sampling method}
\label{sec:covmc}
In  the studies of jet production in $e^+e^-$ 
collisions at LEP~\cite{Heister:2003aj,OPAL:2011aa}, 
PETRA~\cite{Schieck:2012mp} and TRISTAN~\cite{Zomorrodian:2015dom} the 
correlations between measured jet rates were estimated from the 
sampling of MC simulated events. 
 By choosing large number of events out of the set of all  MC simulated 
 events,  $N_{\rm subsamples}$ are built. 
 Then the jet rates are measured at particle level in every subsample
 and  the covariance matrix is estimated as 
\begin{equation}
V^{R}_{ij}=\frac{ \sum_{k=1\dots N_{\rm subsamples} } (R_{i,k}-
\bar{R_{i}})(R_{j,k}-\bar{R_{j}}) }{N_{\rm subsamples}-1},
\end{equation}
where $k$ enumerates measurements obtained in the $k$-th subsample, 
$i$ and $j$ indexes correspond to  Eq.~\ref{YR} and bar denotes the 
mean of quantity over the subsamples.
As this method was often used in the studies of jet production in
 $e^+e^-$ collisions, it is compared  to the classes-based method using 
 the relations to jet rates for the exclusive $k_{T}$ algorithm from 
Sec.~\ref{sec:class:jr}, i.e.\ with the same setup as used to produce 
jet rates in Fig~\ref{fig:rates}.
\subsection{Comparison of classes-based and sampling methods} 
 In this study the sampling method is applied to 
 the  $e^+e^-$  events simulated with the \eesherpa MC 
 program~\cite{Gleisberg:2008ta} and clustered with the exclusive $k_T$ 
 jet algorithm. In total $\NSUBSAMPLES$ subsamples with $\NSTEP$ events 
 each are generated. The used steering card is shown in 
 Listing~\ref{lst:eesherpa}. The correlation matrix, 
 $W^R_{ij}=V^R_{ij}/\sqrt{V^R_{ii}V^R_{j j}}$,  
 calculated with the sampling method is given in 
 Tab.~\ref{tab:correlationsmc012}.
 \setlength{\tabcolsep}{1pt}
\renewcommand{\arraystretch}{1.25}
\begin{table}[!htbp]\centering\tabfontsize
\begin{tabular}{|cccc|cccc|cccc||cc|}\hline
\multicolumn{4}{|c}{$R_2$}&\multicolumn{4}{|c}{$R_3$}&\multicolumn{4}{|c||}{$R_4$}& &\\
{\tiny(0.0010)}&{\tiny(0.0046)}&{\tiny(0.0215)}&{\tiny(0.1000)}&{\tiny(0.0010)}&{\tiny(0.0046)}&{\tiny(0.0215)}&{\tiny(0.1000)}&{\tiny(0.0010)}&{\tiny(0.0046)}&{\tiny(0.0215)}&{\tiny(0.1000)}&&\\\hline\hline
+1.00 &+0.38 &+0.21 &+0.07 &-0.23 &-0.31 &-0.19 &-0.07 &-0.46 &-0.11 &-0.05 &-0.01 &{\tiny(0.0010)}&\multirow{4}{*}{$R_2$}\\
    &+1.00 &+0.50 &+0.14 &+0.23 &-0.78 &-0.48 &-0.14 &-0.47 &-0.35 &-0.10 &-0.06 &{\tiny(0.0046)}&\\
    &    &+1.00 &+0.39 &+0.20 &-0.23 &-0.96 &-0.39 &-0.32 &-0.42 &-0.19 &-0.05 &{\tiny(0.0215)}&\\
    &    &    &+1.00 &+0.10 &+0.04 &-0.35 &-1.00 &-0.14 &-0.27 &-0.16 &-0.05 &{\tiny(0.1000)}&\\\hline
    &    &    &    &+1.00 &-0.07 &-0.18 &-0.10 &-0.76 &-0.25 &-0.07 &-0.01 &{\tiny(0.0010)}&\multirow{4}{*}{$R_3$}\\
    &    &    &    &    &+1.00 &+0.26 &-0.04 &+0.28 &-0.31 &-0.08 &+0.04 &{\tiny(0.0046)}&\\
    &    &    &    &    &    &+1.00 &+0.35 &+0.30 &+0.35 &-0.10 &+0.04 &{\tiny(0.0215)}&\\
    &    &    &    &    &    &    &+1.00 &+0.14 &+0.26 &+0.16 &+0.03 &{\tiny(0.1000)}&\\\hline
    &    &    &    &    &    &    &    &+1.00 &+0.31 &+0.10 &+0.02 &{\tiny(0.0010)}&\multirow{4}{*}{$R_4$}\\
    &    &    &    &    &    &    &    &    &+1.00 &+0.28 &+0.03 &{\tiny(0.0046)}&\\
    &    &    &    &    &    &    &    &    &    &+1.00 &+0.02 &{\tiny(0.0215)}&\\
    &    &    &    &    &    &    &    &    &    &    &+1.00 &{\tiny(0.1000)}&\\\hline
\end{tabular}

\caption{The   correlation matrix for $e^+e^-$ jet rates $R_{2}$, $R_{3}$ and $R_{4}$ as functions of cut parameter (given in brackets) 
calculated  with sampling method. The sample is simulated with 
 $\alpha_s(M_Z)=0.12$. 
}
\label{tab:correlationsmc012}
\end{table}

The confidence intervals of the obtained correlation coefficients 
$\rho=W^{R}_{ij}$ can be calculated using Fisher's 
$z$-transformation~\cite{Fisher:1921}. According to 
Ref.~\cite{Fisher:1921} the variable $z=arctanh(\rho)$ is distributed 
normally with standard deviation of $1/\sqrt{N_{\rm subsamples}-3}$, 
hereby, for $N_{\rm subsamples}=\NSUBSAMPLES$ the $68\%$ confidence 
intervals are  
\FISHER

The correlation matrix  calculated with Eq.~\ref{eq:cov} from the same 
sample is given in Tab.~\ref{tab:correlationsmy012}.
\setlength{\tabcolsep}{1pt}
\renewcommand{\arraystretch}{1.25}
\begin{table}[!htbp]\centering\tabfontsize
\begin{tabular}{|cccc|cccc|cccc||cc|}\hline
\multicolumn{4}{|c}{$R_2$}&\multicolumn{4}{|c}{$R_3$}&\multicolumn{4}{|c||}{$R_4$}& &\\
{\tiny(0.0010)}&{\tiny(0.0046)}&{\tiny(0.0215)}&{\tiny(0.1000)}&{\tiny(0.0010)}&{\tiny(0.0046)}&{\tiny(0.0215)}&{\tiny(0.1000)}&{\tiny(0.0010)}&{\tiny(0.0046)}&{\tiny(0.0215)}&{\tiny(0.1000)}&&\\\hline\hline
+1.00 &+0.38 &+0.20 &+0.08 &-0.25 &-0.29 &-0.19 &-0.08 &-0.44 &-0.15 &-0.05 &-0.00 &{\tiny(0.0010)}&\multirow{4}{*}{$R_2$}\\
    &+1.00 &+0.53 &+0.22 &+0.28 &-0.77 &-0.50 &-0.22 &-0.51 &-0.38 &-0.12 &-0.00 &{\tiny(0.0046)}&\\
    &    &+1.00 &+0.42 &+0.24 &-0.23 &-0.96 &-0.42 &-0.36 &-0.45 &-0.23 &-0.01 &{\tiny(0.0215)}&\\
    &    &    &+1.00 &+0.12 &-0.04 &-0.38 &-1.00 &-0.17 &-0.27 &-0.19 &-0.02 &{\tiny(0.1000)}&\\\hline
    &    &    &    &+1.00 &-0.11 &-0.22 &-0.12 &-0.76 &-0.25 &-0.08 &-0.00 &{\tiny(0.0010)}&\multirow{4}{*}{$R_3$}\\
    &    &    &    &    &+1.00 &+0.27 &+0.04 &+0.30 &-0.30 &-0.09 &-0.00 &{\tiny(0.0046)}&\\
    &    &    &    &    &    &+1.00 &+0.38 &+0.33 &+0.37 &-0.06 &-0.00 &{\tiny(0.0215)}&\\
    &    &    &    &    &    &    &+1.00 &+0.17 &+0.27 &+0.19 &-0.00 &{\tiny(0.1000)}&\\\hline
    &    &    &    &    &    &    &    &+1.00 &+0.33 &+0.10 &+0.00 &{\tiny(0.0010)}&\multirow{4}{*}{$R_4$}\\
    &    &    &    &    &    &    &    &    &+1.00 &+0.32 &+0.01 &{\tiny(0.0046)}&\\
    &    &    &    &    &    &    &    &    &    &+1.00 &+0.04 &{\tiny(0.0215)}&\\
    &    &    &    &    &    &    &    &    &    &    &+1.00 &{\tiny(0.1000)}&\\\hline
\end{tabular}

\caption{The   correlation matrix for $e^+e^-$ jet rates  $R_{2}$, $R_{3}$ and $R_{4}$  as functions of cut parameter (given in brackets) 
calculated with classes-based method. The sample is simulated with 
 $\alpha_s(M_Z)=0.12$. 
}
\label{tab:correlationsmy012}
\end{table}
The classes-based method and the sampling method with 
large number of subsamples give very close results. However, the results 
obtained with  sampling method have sizeable uncertainties, especially 
for the low values of correlation coefficients. 
This property of the sampling method comes from the ignoring 
 addition information of event independence within the subsamples.

\begin{figure}[!htbp]\centering
\resizebox{1.0\linewidth}{0.75\linewidth}{
{\bf \boldmath\input{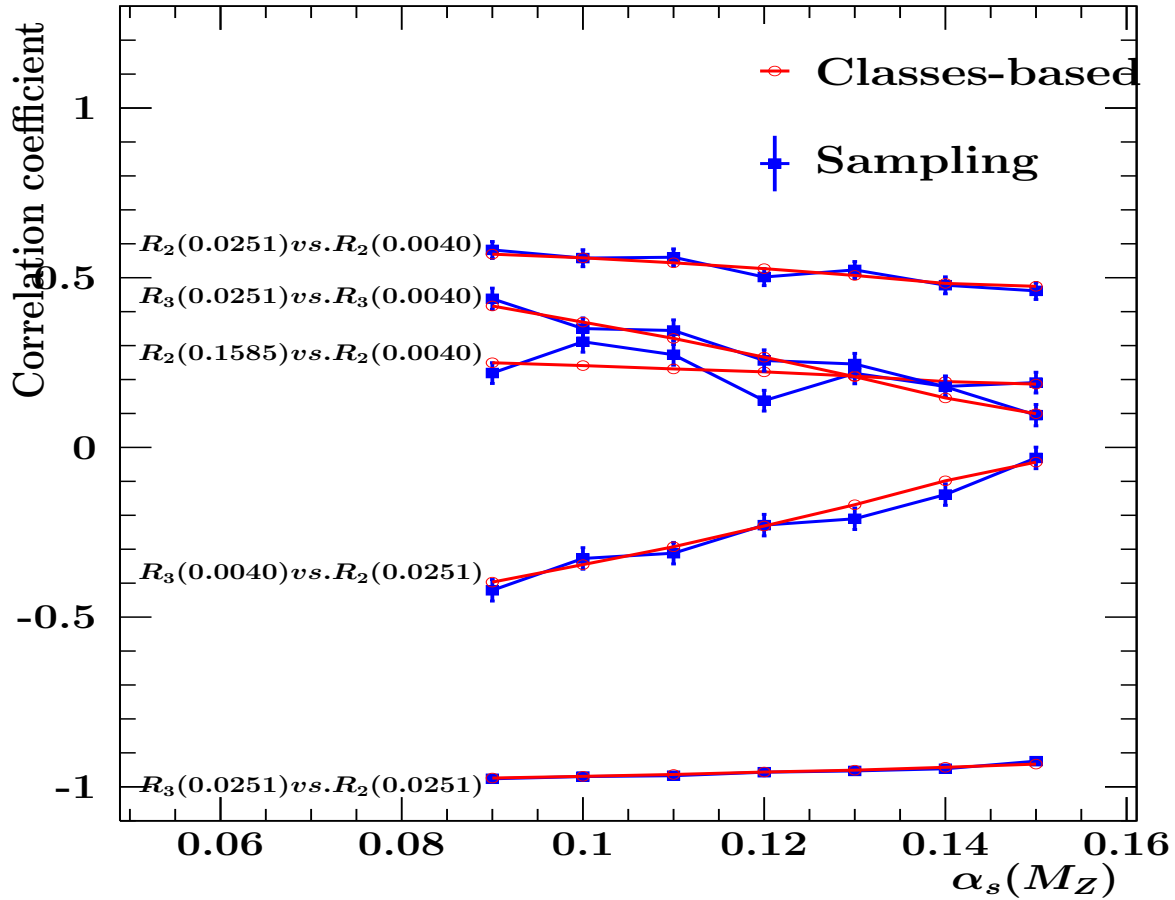}}
}
\caption{Dependence of selected elements of correlation matrix for jet rates $W^{R}$ 
calculated with sampling method and with classes-based 
method
on $\alpha_s(M_Z)$ used by \eesherpa in simulation.
 The error bars for the sampling method are the $68\%$ confidence 
intervals calculated according to 
Ref.~\protect\cite{Fisher:1921}.
 For better visibility the individual points are connected with straight lines.
}
\label{fig:Veesherpa}
\end{figure}

The stability of classes-based and sampling methods can be also tested 
and compared in another way.
In perturbative QCD jet rates at every $y$ cut value are smooth  functions of 
$\alpha_s(M_Z)$~\cite{Weinzierl:2010cw}.
The correlation coefficients depend on the jet rates, therefore, in 
case of proper estimation of correlation coefficients,  
a smooth dependence of correlation coefficients on the  
$\alpha_s(M_Z)$ is expected.
To study dependence of the obtained results on $\alpha_s(M_Z)$, 
the calculations from previous sections are repeated with samples 
 generated using  $\alpha_s(M_Z)=0.09,0.10,0.11,$\arxivbreak{}
 $0.12,0.13,0.14,0.15$. 
 Some of the obtained results are shown in Fig.~\ref{fig:Veesherpa}.

 Smooth dependency of correlation coefficients obtained with classes-based 
 method on $\alpha_s(M_Z)$ can serve as a proof of method robustness.
 For both methods  significant dependence  on the value of  $\alpha_s(M_Z)$ 
 used in the simulation  is present. 
Therefore, to obtain  correlation coefficients consistent with data,
sampling of data  or classes-based method  should be used. While for 
the former option the precision of the obtained correlation coefficients 
 is limited by number of subsamples~\cite{Fisher:1921}, the 
 classes-based method does not face this problem.
 In addition, the 
 sampling of data involves complications with corrections of the 
 measured quantities for detector effects. Contrary to that, the 
 correction for detector effects for classes is  straightforward, can 
 be done either ``bin-by-bin'', or with the unfolding  
 procedures and propagated to the 
 final results.
\FloatBarrier
\section{Summary}
\label{sec:conc}
A new type of jet observables, classes, is introduced. 
The classes-based method to calculate correlations between measurements 
of jet observables is demonstrated. The  method  provides robust 
results, does not rely on the MC simulations and has a straightforward, 
self-consistent procedure for taking into account detector corrections.
\section*{Acknowledgements}
\label{sec:ack}
I thank Stefan Kluth for his major contribution to the development 
of this paper. I thank Olaf Behnke and Oleksandr Zenaiev for  the 
discussions on the topic and suggestions for the improvements of the 
manuscript.
\begin{appendices}
\section{Example of \eesherpa setup for $e^{+}e^{-}$ collision simulation}
\label{sec:setup}
\begin{lstlisting}[caption=\eesherpa steering card for $e^+e^-$ sample 
simulated with {$\alpha_s(M_Z)=0.12$}.,language=bash,label=lst:eesherpa]
(run){
  NJET:=2;
  ALPHAS(MZ) 0.12;
  ORDER_ALPHAS 2;
  BEAM_1  11; BEAM_ENERGY_1 45.6;
  BEAM_2 -11; BEAM_ENERGY_2 45.6;
}(run)
(isr){
  PDF_LIBRARY None;
}(isr)
(processes){
  Process 11 -11 -> 93 93 93{NJET};
  CKKW pow(10,-2.25);
  Order (*,2);
  End process;
}(processes)
\end{lstlisting}
\section{Example of \ppsherpa setup for $pp$ collision simulation}
\label{sec:ppsetup}
\begin{lstlisting}[caption=\ppsherpa steering card for $pp$ sample.,language=bash,label=lst:ppsherpa]
(run){
  BEAM_1 2212; BEAM_ENERGY_1 3500;
  BEAM_2 2212; BEAM_ENERGY_2 3500;
  ALPHAS(MZ) 0.12;
}(run)
(isr){
  PDF_LIBRARY=LHAPDFSherpa
  PDF_SET=HERAPDF20_NNLO_ALPHAS_120
}(isr)
(processes){
  Process 93 93 -> 93 93 93{0};
  Order (*,0);
  CKKW sqr(20/E_CMS)
  Integration_Error 0.05;
  End process;
}(processes)
(selector){
  NJetFinder  2  50.0  0.0  0.4  -1 3.0
}(selector)
\end{lstlisting}
\section{Example of \epsherpa setup for $e^{\pm}p$ collision simulation}
\label{sec:epsetup}
\begin{lstlisting}[caption=\epsherpa steering card for $e^{+}p$ sample.,language=bash,label=lst:epsherpa]
(run){
  NJET:=2; QCUT:=5; SDIS:=0.6;  LJET:=2,3; LGEN:=BlackHat;
  ME_SIGNAL_GENERATOR Comix Amegic LGEN;
  EVENT_GENERATION_MODE Weighted;
  RESPECT_MASSIVE_FLAG 1;  CSS_KIN_SCHEME 1;
  ALPHAS(MZ) 0.12;
  BEAM_1 2212 920; BEAM_2 11 27.5;
}(run)
(isr){
  PDF_LIBRARY=LHAPDFSherpa
  PDF_SET_1=HERAPDF20_NNLO_ALPHAS_120
  PDF_SET_2 None;
}(isr)
(processes){
  Process  93 11-> 11 93 93{NJET};
  CKKW sqr(QCUT/E_CMS)/(1.0+sqr(QCUT/SDIS)/Abs2(p[2]-p[1]));
  NLO_QCD_Mode MC@NLO {LJET};
  Order (*,2); Max_N_Quarks 6;
  ME_Generator Amegic {LJET};
  RS_ME_Generator Comix {LJET};
  Loop_Generator LGEN;
  PSI_ItMin 25000 {3};
  Integration_Error 0.05 {3};
  End process;
}(processes)
(selector){
  Q2 11 11 120 6000
}(selector)
\end{lstlisting}
\end{appendices}

{\bibliographystyle{./JetClasses}{
\raggedright\bibliography{JetClasses.bib}
}
\vfill\eject
\end{document}